\documentclass[aps,twocolumn,showpacs,preprintnumbers,nofootinbib,tightenlines,useAMS]{revtex4}
\usepackage{bm}
\usepackage{times}
\usepackage[dvips]{graphicx}
\usepackage{subfigure}
\usepackage{caption}
\begin{document}

\title{The nuclear pseudospin symmetry along an isotopic chain}

\author{Ronai Lisboa, Manuel Malheiro}

\affiliation{Instituto de F\'{\i}sica, Universidade Federal
Fluminense, Nit\'eroi 24210-340, RJ, Brazil}

\author{P. Alberto}

\affiliation{Departamento de F{\'\i}sica and Centro de F{\'\i}sica
Computacional, Universidade de Coimbra, P-3004-516 Coimbra
Portugal}
\begin{abstract}
{We investigate the isospin dependence of pseudospin symmetry in
the chain of tin isotopes (from $^{120}$Sn until $^{170}$Sn).
Using a Woods-Saxon parametrization of the nuclear potential for
these isotopes we study in detail the effect of the
vector-isovector $\rho$ and Coulomb potentials in the energy
splittings of neutron and proton pseudospin partners in the
isotopic chain. We conclude that the realization of nuclear
pseudospin symmetry does not change considerably with the mass
number, and is always favored for neutrons. We also find that the
$\rho$ potential accounts for essentially all the pseudospin
isospin asymmetry observed and that the Coulomb potential plays a
negligible role in this asymmetry. This can be explained by the
dynamical nature of pseudospin symmetry in nuclei, namely the
dependence of the pseudospin splittings on the shape of the
nuclear mean-field potential.}
\end{abstract}
\maketitle
\section{Introduction}

 Pseudospin was introduced in the late 60's \cite{kth,aa}
to account for the quasi-degeneracy of single-nucleon states with
quantum numbers ($n$, $\ell$, $j= \ell + 1/2$) and ($n-1$,
$\ell+2$, $j= \ell$ + 3/2) where $n$, $\ell$, and $j$ are the
radial, the orbital, and the total angular momentum quantum
numbers, respectively. These levels have the same ``pseudo''
orbital angular momentum quantum number, $\tilde{\ell} = \ell +
1$, and ``pseudo'' spin quantum number, $\tilde s = 1/2$.

 The pseudospin symmetry has been analyzed in the framework
 of non-relativistic models
by several authors \cite{bahri,mosh,mosk}. The subject was revived
in recent years when Ginocchio \cite{gino} presented a
relativistic interpretation for this symmetry showing that the
quantum number $\tilde\ell$ is the orbital quantum number of the
lower component of the Dirac spinor for spherical potentials.
Moreover, he showed that $\tilde\ell$ is a good quantum number in
a relativistic theory for the nucleon with scalar $S$ and vector
$V$ potentials with opposite signs and the same magnitude, i.e.,
$\Sigma=S+V=0$. Actually, this condition can be relaxed to demand
that only the derivative of $\Sigma$ be zero \cite{arima}, but, if
$\Sigma$ goes to zero at infinity, both conditions are equivalent.

 Unfortunately, neither of these two conditions are met in nuclei.
Some recent works indicate that the nuclear pseudospin symmetry
has a dynamical character. It was shown that the almost degeneracy
of the pseudopsin doublets not only depends in the shape of the
nuclear mean fields \cite{pmmdm_prl} but in fact arises from a
significant cancellation among the terms that contribute to the
energy and not only from the pseudospin orbit interaction
\cite{pmmdm_prc,H2002}. The non-perturbative nature of this
interaction has been discussed in \cite{marcos,marcos2}, where it
also was pointed out the dynamical character of the nuclear
pseudospin.

 We have investigated recently in a quantitative way the effect
of the Coulomb and the vector-isovector $\rho$ potentials in the
proton and neutron asymmetry seen in the nuclear pseudospin
\cite{ronai,prc67}. We analyzed the effect of these potentials in
each of the terms that contribute to the pseudospin energy
splitting. We concluded that the isospin asymmetry seen in the
nuclear pseudospin is also a manifestation of the dynamical
character of this symmetry.

 In this paper we will present new results concerning the isospin
dependence of the nuclear pseudospin along Sn isotopic chain. We
will plot the energy splitting for some pseudospin doublets as a
function of the mass number $A$, analyzing separately the neutron
and proton levels. To this end, we solve the Dirac equation for
the nucleons using a parametrization for the nuclear potential
along this chain which was developed by us in a recent work
\cite{prc67}. That parametrization was obtained by a fit to a
sophisticated self-consistent relativistic mean field calculation
\cite{meng}. We will show that the pseudospin quasi-degeneracy has
a small dependence with $A$ in accordance with \cite{meng0}.

 The symmetry is better realized for neutrons
than is for protons. Furthermore, as found in previous works
\cite{pmmdm_prl,prc67}, we also show that the main reason for the
difference between the energy splitting of proton and neutron
pseudospin partners comes from the vector-isovector potential and
not from the Coulomb potential.

 The paper is organized as follows: in section II we present the
nuclear potentials as combinations of Woods-Saxon potentials and
the respective parameters as functions of $A$ and $N-Z$. In
section III we analyze the energy splittings of two pseudospin
doublets that are close to the Fermi sea, for neutrons and
protons, as a function of $A$ from $120$ to $170$. In particular,
we look for the effect of the Coulomb and vector-isovector
potentials in those energy splittings. Finally, we draw our
conclusions.
\vspace*{-.5cm}
\section{Dirac equation and pseudospin symmetry}

 The Dirac equation for a particle of
mass $m$ in external scalar $S$ and vector $V$ potentials is given
by
\begin{equation}
H\Psi=[\bm{\alpha}\cdot\bm{p}\,+\,\beta(m\,+\,S)\,+\,V]\Psi\,=\,\epsilon\,\Psi\,,\label{dirac}
\end{equation}
where $\bm\alpha$ and $\beta$ are the usual Dirac matrices. The
Hamiltonian in Eq.~(\ref{dirac}) is invariant under SU(2)
transformations when $S = V$ or $S = - V$ \cite{smith,bell,levi}.
The second case corresponds to the realization of pseudospin
symmetry. As referred before, this symmetry is related to the
orbital quantum number of the lower component of the Dirac spinor.

If $S$ and $V$ are radial potentials, equation (\ref{dirac}) gives
rise to two differential equations for the upper and lower radial
wave functions. Defining $\Delta=V-S$, $\Sigma=V+S$, and the
binding energy $E=\epsilon-m$, these are
\begin{eqnarray}
\frac{1}{r^2}\frac{d\hfill}{d\,r}\bigg(r^2\,\frac{d\,G_\kappa}{d\,r}\bigg)-
\frac{\ell(\ell+1)}{r^2}G_\kappa+\frac{\Delta'}{E+2m-\Delta}\nonumber\\\bigg(G'_\kappa+
\frac{1+\kappa}{r}G_\kappa\bigg)+
(E+2m-\Delta)(E-\Sigma)G_\kappa=0\\\label{upper}
\frac{1}{r^2}\frac{d\hfill}{d\,r}\bigg(r^2\,\frac{d\,F_\kappa}{d\,r}\bigg)-
\frac{\tilde\ell(\tilde\ell+1)}{r^2}F_\kappa+\frac{\Sigma'}{E-\Sigma}\nonumber\\\bigg(F'_\kappa
+\frac{1-\kappa}{r}F_\kappa\bigg)+(E+2m-\Delta)(E-\Sigma)F_\kappa=0\,,\label{lower}
\end{eqnarray}
where $\kappa$ is the quantum number defined by
\begin{equation}
\kappa=\left\{\begin{array}{cl}
                    -(\ell+1)         &\quad\,j   =  \ell + \frac{1}{2}\\
                      \ell            &\quad\,j   =  \ell - \frac{1}{2}
                   \end{array}\right.\,.
\end{equation}
The term with $1-\kappa$ in Eq.~(\ref{lower}) is the
pseudospin-orbit term \cite{palberto}. From that equation one sees
that, should it be possible to set $\Sigma'=0$, $\tilde\ell$ would
be a good quantum number. Since the sign of $\kappa$ determines
whether one has parallel or antiparallel spin and
$\tilde\ell=\ell-\kappa/|\kappa|$, one sees that pairs of states
with $\kappa=-(\ell+1)$ and $\kappa=\ell+2$ have the same
$\tilde\ell=\ell+1$, the quantum numbers of the pseudospin
partners. For example, for $[ns_{1/2},(n-1)d_{3/2}]$ one has
$\tilde{l}=1$, for $[np_{3/2},(n-1)f_{5/2}]$ onde has
$\tilde{l}=2$ etc. Pseudospin symmetry is exact when doublets with
$j=\tilde{l}\pm\tilde{s}$ are degenerate. \vspace*{-.4cm}
\section{THE NUCLEAR POTENTIAL FOR THE {Sn} ISOTOPIC CHAIN}

In a recent work we found a general parametrization in terms of
Woods-Saxon potentials for the binding potential $\Sigma$ in the
whole Sn isotopic chain. The procedure used to extract the
parameters is explained in detail in \cite{prc67}. We separated
$\Sigma$ in a central $\Sigma_c(r)$, a vector-isovector
$V_\rho(r)$ and a Coulomb part $V_{\rm Coul}(r)$ (only for
protons), in the following way
\begin{eqnarray}
\Sigma(r)_{p,n}&=&\frac{\Sigma_{o\,c}}
{1+\exp[(r-R_{c})/a_{c}]}\pm\frac{V_{o\,\rho}}
{ 1+\exp[(r-R_{\rho})/a_{\rho}]}\nonumber\\
&&+V_{\rm Coul}(r)\ . \label{potfull}
\end{eqnarray}
In $V_\rho$ the plus sign is for protons and the minus sign for
neutrons. The parameters, as functions of $A$, $N$ and $Z$, are
\begin{eqnarray}
\Sigma_{o\,c}&=&-69.94\ {\rm MeV}\\
R_{c}&=&1.21A^{1/3}\ {\rm fm}\\
a_{c}&=&0.13A^{1/3}\ {\rm fm}\\
V_{o\,\rho}&=&-[ 0.12(N-Z)+3.87]\ {\rm MeV}\label{v0rho}\\
R_{\rho}&=&[0.03(N-Z)+5.05]\ {\rm fm}\label{rrho}\\
a_{\rho}&=&[0.007(N-Z)+0.27]\ {\rm fm}\ .\label{arho}
\end{eqnarray}

 The parametrization for $R_c$ and $a_c$ has a natural
justification in view of the known $A^{1/3}$ dependence of the
nuclear radius. The $V_{o\,\rho}$, $R_\rho$ and $a_\rho$
dependencies on $N-Z$ are also justified since they are
proportional to the difference between proton and neutron
densities. For $V_{\rm Coul}$ we take the proton electrostatic
potential energy in a uniform spherical charge distribution of
charge $Ze$ and radius $R_c$
\begin{eqnarray}
\label{VCoul} V_{\rm Coul}(r)=\left\{\begin{array}{ll}
            \frac{1}{4\pi\epsilon_{0}}\frac{Ze^2}{2R_{C}}
\biggl(3-\frac{r^2}{R^2_{C}}\biggr)&,\quad\;r<\;R_{C}\\[0.3cm]
            \frac{1}{4\pi\epsilon_{0}}\frac{Ze^2}{r}&,\quad\;r\geq\;R_{C}
            \end{array}
       \right.\quad .
\end{eqnarray}
We used in Eq.~(\ref{VCoul}) $R_{C}=1.20A^{1/3}$ which has the
same $A$ dependence of the nuclear radius.

\begin{figure}[hbt]
\parbox[t]{6.cm}{
\begin{center}
\includegraphics[width=6.5cm,height=6.5cm]{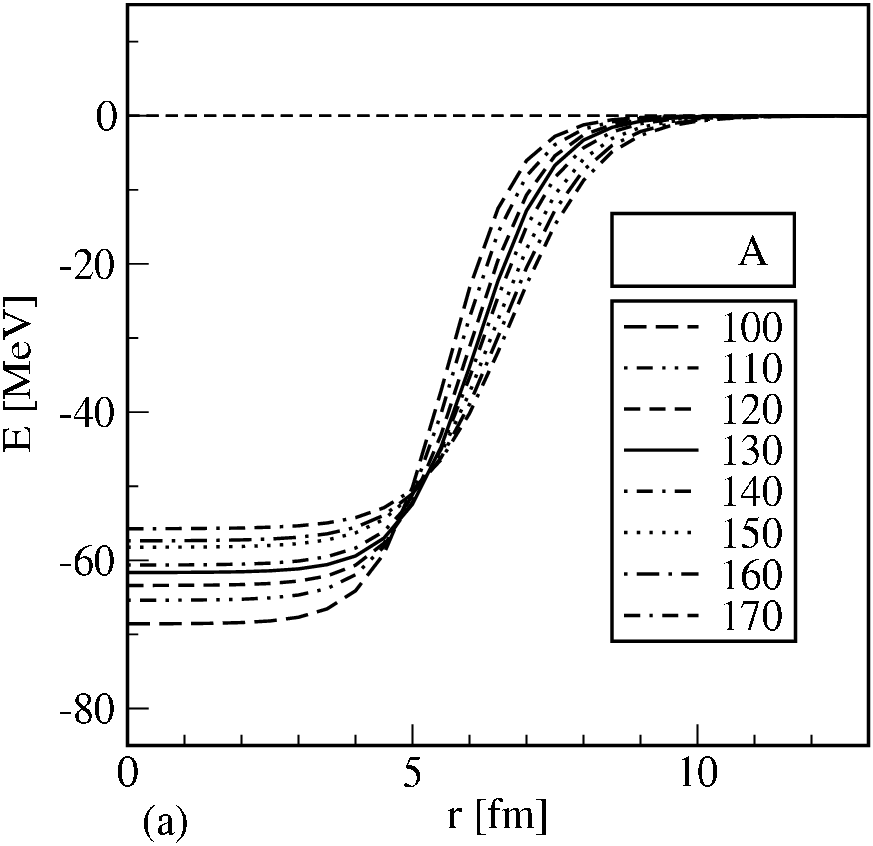}
\end{center}\vspace*{-0.4cm}
}\hfill
\parbox[t]{6.cm}{
\begin{center}
\includegraphics[width=6.5cm,height=6.5cm]{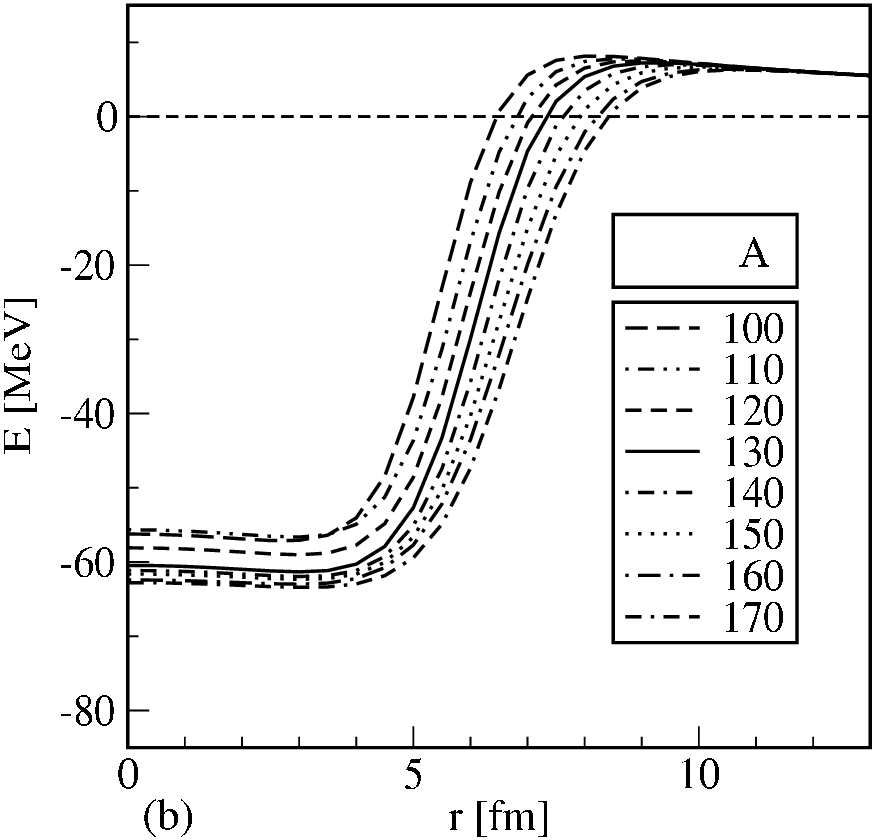}
\end{center}\vspace*{-0.4cm}
}%
\caption{Woods-Saxon potential for neutrons (a) and protons (b)
along the Sn isotopic chain.\hspace*{3cm plus 1cm minus
1cm}}\label{fig:fit}
\end{figure}

 We show in Fig.~\ref{fig:fit} the binding $\Sigma$ potential
for neutrons and protons in the Sn isotopic chain. In the proton
case we see explicitly the Coulomb barrier produced by the
potential (\ref{VCoul}).
\section{ISOSPIN DEPENDENCE OF PSEUDOSPIN SYMMETRY}

 Using the nuclear potential presented in Eq.~(\ref{potfull}) we now display
the energy splitting for the two pseudospin doublets $[3s_{1/2} -
2d_{3/2}]$ and $[2d_{5/2} - 1g_{7/2}]$. Firstly we present results
for the neutrons and after for the protons where we need to
consider also the Coulomb potential.
\begin{figure}[hbt]
\parbox[!t]{8cm}{
\includegraphics[width=7cm,height=7cm]{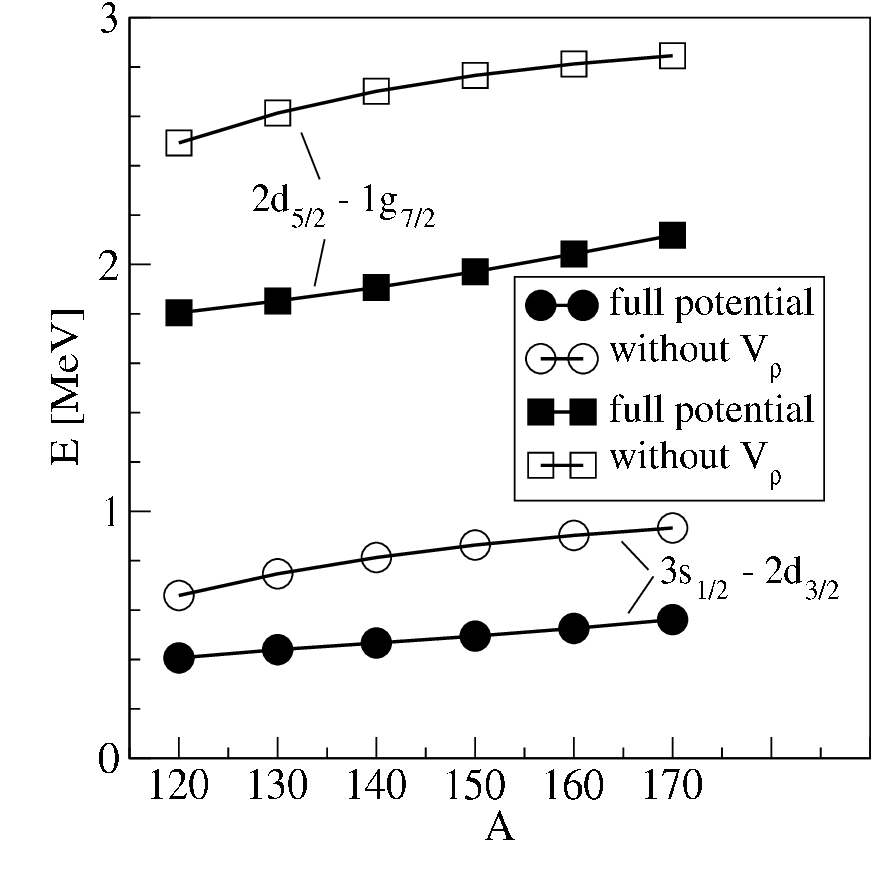}
\vspace*{-0.5cm} \caption{Energy splittings of the pseudospin
pairs $[3s_{1/2} - 2d_{3/2}]$ and $[2d_{5/2} - 1g_{7/2}]$ as a
function of the mass number $A$ for neutrons with and without
$V_{\rho}$.\hspace*{2cm minus 1cm}}\label{fig:neu}}
\end{figure}

 In Fig.~\ref{fig:neu} is shown the neutron energy splittings for
those pseudospin partners as $A$ changes. In both cases the energy
differences increase just slightly as $A$ increases. Since the
pair $[2d_{5/2} - 1g_{7/2}]$ is more bound than $[3s_{1/2} -
2d_{3/2}]$, one may notice also that the energy splittings are
smaller for levels closer to the the Fermi sea.

In a recent work \cite{pmmdm_prl} a systematics was established
relating the pseudospin splittings to the Woods-Saxon parameters
of the nuclear potentials. The conclusions were that the energy
differences become smaller as the magnitude of the well depth
$|\Sigma|$ decreases and as the diffusivity (of both $\Sigma$ and
$\Delta$ potentials) increases, but grow when their radius
increases. From Fig.~\ref{fig:fit} we see that in the isotope
chain the magnitude of the neutron potential well decreases and
its diffusivity increases as $A$ increases, thus favoring the
pseudospin symmetry. However, the increase of the radius with
$A^{1/3}$ works in the opposite way and thus essentially cancels
the previous effects. Thus, the dependence with $A$ of pseudospin
splittings for neutrons is small.
\begin{figure}[hbt]
\includegraphics[width=7cm,height=7cm]{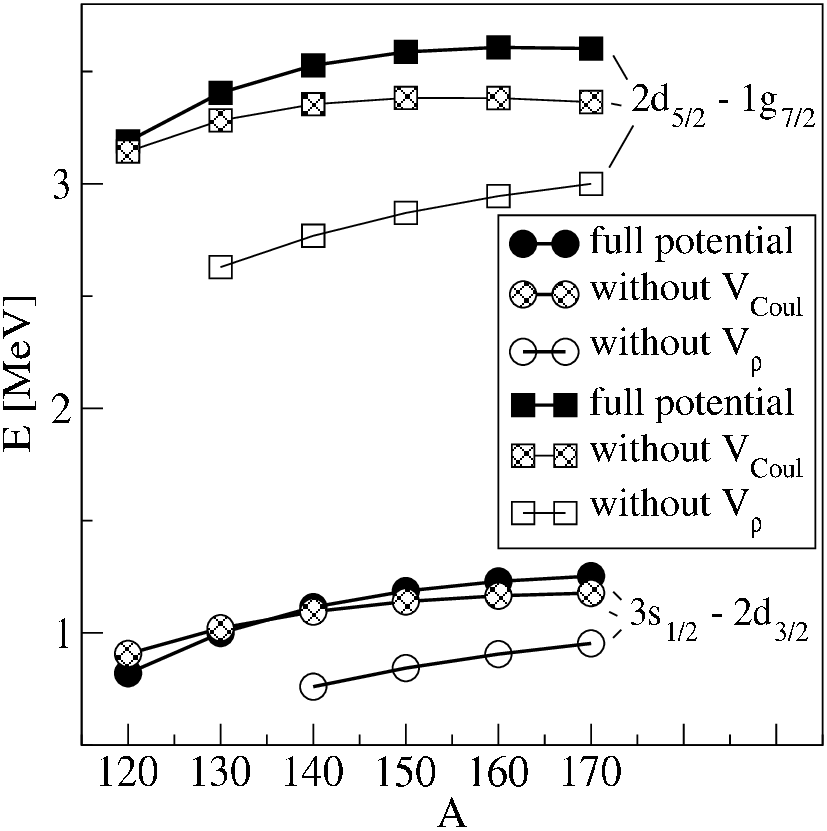}
\vspace*{-0.5cm} \caption{Energy splittings of the pseudospin
pairs $[3s_{1/2} - 2d_{3/2}]$ and $[2d_{5/2} - 1g_{7/2}]$ as a
function of the mass number $A$ for protons with and without
$V_{\rho}$ and $V_{\rm Coul}$.\hspace*{2cm minus
1cm}}\label{fig:pro}
\end{figure}

 We now perform the same study for protons. The result is presented
in Fig.~\ref{fig:pro}. Again, we see that there is almost no
change of the energy splittings with the mass number $A$. Also, as
in the neutron case, the splittings are smaller for higher levels.
For example, for $A$=$150$ the energy difference for the doublet
$[3s_{1/2},2d_{3/2}]$ is $1.19$ MeV for protons and is $0.49$ MeV
for neutrons, whereas the $[2d_{5/2},1g_{7/2}]$ pair has an energy
difference of $3.58$ MeV for protons and of $1.97$ MeV for
neutrons. We see that, for the same value of $A$, pseudospin
symmetry is favored for neutrons.

What is the origin of this experimentally observed isospin
asymmetry of pseudospin in nuclei? Clearly this must be related to
both the $V_{\rho}$ potential, which is repulsive for neutrons and
attractive for protons, and the Coulomb potential, a repulsive
potential which only exists for protons.

Since the Coulomb potential only affects protons, we could think
that it could give an important contribution for this asymmetry.
However, from Fig.~\ref{fig:pro} we see that, when the Coulomb
potential is turned off, the pseudospin splittings remain
practically the same. For example, for $A$=$150$, if we remove the
Coulomb potential, the difference of energy for the doublet
$[3s_{1/2},2d_{3/2}]$ is $1.13$ MeV (compare with the value of
1.19 MeV given before), whereas for $[2d_{5/2},1g_{7/2}]$ is
$3.38$ MeV ($3.58$ MeV).

 In \cite{prc67} we saw that this fact is due to a
cancellation between the diverse terms that contribute to the
energy of each level. In that study, we showed that, while the
contribution of the Coulomb potential to the pseudospin-orbit term
is substantial, it is cancelled by the contribution of all the
other terms. Fig.~\ref{fig:pro} shows that, although small, the
Coulomb potential contribution is bigger for the pseudospin
splitting of deeper pseudospin partners.

 In a similar way, we analyze the role of the potential
$V_{\rho}$ in the isospin asymmetry by looking into its effect in
the splittings of the neutron and proton pseudospin pairs. In
Fig.~\ref{fig:neu} is presented the neutron energy difference of
these pairs when the potential $V_{\rho}$ is excluded. When this
is done, the energy difference increases, showing that this
potential favors pseudospin symmetry for neutrons, which can be
understood by the systematics referred to before, since
$V_{\rho}$, being positive, decreases $|\Sigma|$ (note that
$\Sigma<0$), and at the same time makes it more diffuse. This can
be see in Fig.~\ref{fig:fit} for the neutrons: $V_{\rho}$ is
larger as $A$ and $N-Z$ increases and, as a consequence, the well
potential is less profound and more diffuse. For example, again
for $A$=$150$, the energy difference for the neutron doublet
$[3s_{1/2},2d_{3/2}]$ without the $\rho$ potential is $0.86$ MeV
($0.49$ MeV with the full potential), whereas for
$[2d_{5/2},1g_{7/2}]$ is $2.76$ MeV ($1.97$ MeV).

 In the proton case, shown in Fig.~\ref{fig:pro} the effect of
$V_{\rho}$ is the opposite, as could be expected, but is smaller
in magnitude than in the neutron case. This time $|\Sigma|$
becomes bigger when $V_{\rho}$ is added and less diffuse thus
working against the realization of the pseudospin symmetry. For
$A$=$150$ the energy difference for the doublet
$[3s_{1/2},2d_{3/2}]$ is $0.86$ MeV (previously $1.19$ MeV),
whereas for $[2d_{5/2},1g_{7/2}]$ is $2.87$ MeV ($3.58$ MeV). This
analysis allows us to conclude that when the vector-isovector
potential $V_{\rho}$ is excluded the pseudospin asymmetry for
protons and neutrons almost disappears. Therefore, this potential
is the main responsible for this asymmetry and the Coulomb
potential $V_{\rm Coul}$ does not play a significant role.
\section{Conclusions}

 We have investigated in a quantitative way the isospin dependence of
the nuclear pseudospin along the Sn isotopic chain. To do this
analysis we performed a mean-field model calculation with a
parametrization for the nuclear potential in this chain developed
by us in a recent work \cite{prc67}. We used Woods-Saxon
potentials to fit the proton and neutron potentials obtained by a
sophisticated self-consistent calculation for the Sn isotopes
\cite{meng}.

In this general parametrization of the potential we have separated
explicitly the $V_{\rm Coul}$ and $V_{\rho}$ potentials. These two
potentials are the main source of the isospin dependence of the
nuclear pseudospin. In order to identify the origin of this
dependence we have analyzed in detail the effect of those
potentials separately in the proton and neutron pseudospin energy
splittings along the Sn isotopic chain.

 We conclude that the dependence with mass number $A$ of the
pseudospin symmetry measured by the energy splittings is small
either for neutrons or for protons along the isotopic chain. The
effect of the Coulomb barrier is also very small and almost
negligible for the proton levels close to the Fermi sea. The
difference seen in nature for the pseudospin energy splitting of
the neutrons and protons comes essentially from the
vector-isovector $V_{\rho}$ potential. It makes the binding
$\Sigma$ potential more diffuse for neutrons than for protons.
Thus the energy splitting decreases for neutrons and increases for
protons originating this isospin asymmetry in the nuclear
pseudospin.

 Finally, from our analysis we can conclude that, at least for tin
 isotopes, the realization of
 pseudospin symmetry is almost independent of the neutron content
 of nuclei in a isotopic chain.

\begin{acknowledgments}
R. L. thanks the nice atmosphere during the XV RETINHA where this
work has been presented. P. A. acknowledges the financial support
from FCT (POCTI), Portugal. M. M. acknowledges the financial
support from the CNPq/ICCTI Brazilian-Portuguese scientific
exchange program. M. M and R. L. acknowledge in particular the
CNPq support.
\end{acknowledgments}


\small
\begin{thebibliography}{000}
\bibitem{kth} K. T. Hecht and A. Adler, Nucl. Phys. {\bf A137}, 129 (1969)
\bibitem{aa} A. Arima, M. Harvey, and K. Shimizu, Phys. Lett. {\bf B30},
 517 (1969)
\bibitem{bahri} A. L. Blokhin, C. Bahri, and J. P. Draayer,
Phys. Rev. Lett. {\bf 74}, 4149 (1995)
\bibitem{mosk} C. Bahri,
J. P. Draayer, and S. A. Moszkowski, Phys. Rev. Lett. {\bf 68},
2133 (1992)
\bibitem{mosh} O. Casta\~nos, M. Moshinski, and C. Quesne,
Phys. Lett. {\bf B277} 238 (1992)
\bibitem{gino} J. N. Ginocchio, Phys. Rev. Lett. {\bf 78}, 436
(1997); \textit{ibid}, Phys. Rept. {\bf 315}, 231 (1999)
\bibitem{arima}  J. Meng, K. Sugawara-Tanabe, S. Yamaji,
 P. Ring, and A. Arima, Phys. Rev. {\bf C58}, R628 (1998)
\bibitem{pmmdm_prl}P. Alberto, M. Fiolhais, M. Malheiro, A. Delfino and
M. Chiapparini, Phys. Rev. Lett. {\bf 86}, 5015 (2001)
\bibitem{pmmdm_prc}P. Alberto, M. Fiolhais, M. Malheiro, A. Delfino, and
M. Chiapparini, Phys. Rev. {\bf C65}, 034307 (2002)
\bibitem{H2002} R. Lisboa, M. Malheiro, A. Delfino, P. Alberto,
M. Fiolhais and M. Chiapparini, ``$8$th/interaction Workshop on
Hadrons Physics 2000; nucl-th 0207068''
\bibitem{marcos} S. Marcos, L. N. Savushkin, M. L\'opez-Quelle and
P. Ring, Phys. Rev. {\bf C62}, 054309 (2000)
\bibitem{marcos2} S. Marcos, M. L\'opez-Quelle, R. Niembro, L. N. Savushkin, and
P. Bernardos, Phys. Lett. {\bf B513}, 30 (2001)
\bibitem{ronai} R. Lisboa, ``The dynamical character of the isospin
asymmetry in the nuclear pseudospin'', dissertation (in
portuguese) (Universidade Federal Fluminense, Niter\'oi, 2002)
(unpublished)
\bibitem{prc67} R. Lisboa, M. Malheiro and P. Alberto, Phys. Rev. {\bf C67}, 054305 (2003)
\bibitem{meng} J. Meng and I. Tanihata, Nucl. Phys. {\bf A650},
 176 (1999)
\bibitem{meng0} J. Meng, K. Sugawara-Tanabe, S. Yamaji,
and A. Arima, Phys. Rev. {\bf C59}, 154 (1999)
\bibitem{smith}B. Smith and L. J. Tassie, Ann. Phys. {\bf 65},
 352 (1971)
\bibitem{bell} J. S. Bell and H. Ruegg, Nucl. Phys. {\bf B98},
 151 (1975)
\bibitem{levi} J. N. Ginocchio and A. Levitan, Phys.Lett. {\bf
B245}, 1 (1998)
\bibitem{palberto} P. Alberto, M. Fiolhais, and M. Oliveira,
Eur. J. Phys. {\bf 19}, 553 (1998)
\end{thebibliography}
\end{document}